\documentclass[aps,twocolumn]{revtex4}
\usepackage{latexsym}
\usepackage{amsfonts}
\usepackage{graphicx}
\usepackage{epsfig}

\begin{document}

\title{New class of level statistics in correlated disordered chains}

\author{Pedro Carpena}
\author{Pedro Bernaola-Galv\'an}
\affiliation{Departamento de F\'{\i}sica Aplicada II. E.T.S.I. de
Telecomunicaci\'on, Universidad de M\'alaga. 29071, M\'alaga,
Spain.}
\author{Plamen Ch. Ivanov}
\affiliation{Center for Polymer Studies and Department of Physics,
Boston University, Boston, MA 02215.}

\begin{abstract}
We study the properties of the level statistics of 1D disordered
systems with long-range spatial correlations. We find a threshold
value in the degree of correlations below which in the limit of
large system size the level statistics follows a Poisson
distribution (as expected for 1D uncorrelated disordered systems),
and above which the level statistics is described by a new class
of distribution functions. At the threshold, we find that with
increasing system size the standard deviation of the function
describing the level statistics converges to the standard
deviation of the Poissonian distribution as a power law. Above the
threshold we find that the level statistics is characterized by
different functional forms for different degrees of correlations.
\end{abstract}

\maketitle

Investigating statistical properties of the energy spectrum and
the behavior of the level statistics proved to be a useful
approach to study electronic properties of disordered
systems\cite{alt86,mir00,cue98}. In 3D disordered systems the
Anderson transition between metallic and insulating phase is
associated with a transition in the level statistics distribution
from a Wigner-Dyson form to Poisson distribution
\cite{shk93,hof93,meh90}, and the level statistics at the critical
point has been studied\cite{shk93}. In 1D disordered systems, for any degree of
disorder, the level statistics is described by the Poisson
distribution in the limit of large system size corresponding to
electronic localization and insulating behavior \cite{wis90}.

Recently, it has been demonstrated numerically that introducing
long-range correlations in the spatial order of atoms with
different energies in a chain can lead to electronic
delocalization \cite{mou98}, creating an interesting debate\cite{comment}.
Further, it has been shown that there
is a localization-delocalization transition at a critical value of
the degree of correlations imposed on the disorder in the system,
and later works extended these results to other models
\cite{zha02,izr99,izr01,car02}. Also, this type of transition can be found
in quasiperiodic systems, as the Aubrey-Andre model and other models\cite{var92},
indicating the importance of some type of ordering.
These theoretical findings are
supported also by experimental results showing delocalization and
electronic transport driven by extended states in
correlated-disordered GaAs/Ga$_{0.7}$Al$_{0.3}$ superlattices
\cite{bel99}. Recently, a metal-insulator transition has been
reported in 2D correlated disordered systems \cite{liu03}.

Here we hypothesize that, as the localization properties of the
electronic states in a disordered system are affected by the
degree of spatial correlations\cite{mou98,car02}, the properties
of the level statistics of the energy spectrum of such
correlated disordered systems could also change. The effect of correlations
in the level statistics can be combined with the fact that interactions, which
are extremely important in 1D, also affect the level statistics \cite{bla01}.
Specifically, we
investigate how the functional form of the distribution describing
the level statics is affected by the degree of correlations
introduced in the system. We demonstrate that the Poissonian form
describing the level statistics in disordered 1D systems in the
thermodynamic limit is preserved even when certain degree of
spatial correlations is introduced. Further, we find a critical
threshold for the degree of correlations above which in the
thermodynamic limit there is a transition to a different class of
distribution functions for the level statistics.

We consider the standard 1D tight-binding hamiltonian with
nearest-neighbor interaction
\begin{equation}
H = \sum_i \xi_i \,|i> <i| + \sum_{<i,j>} V \, |i><j|,
\label{hamil}
\end{equation}
where $V$ is the coupling energy and $i$ ranges from $1$
to $N$, where $N$ is the system size. To fix the energy scale we
choose $V=1$. In the case of uncorrelated disorder, the site
energies $\{ \xi_i \}$ are randomly drawn from a certain
probability distribution, commonly a box (uniform) distribution or
a Gaussian. This is equivalent to consider the series of site
energies as white noise. In contrast, for systems with correlated
disorder we introduce spatial long-range correlations in the
series of site energies $\{ \xi_i \}$, so that their
sequence describes the trace of a fractional Brownian motion. To
this end, we obtain the site energies using the inverse Fourier
transform
\begin{equation}
\xi _{i}=\sum_{k=1}^{N/2}\left[ k^{-\beta }\left( \frac{2\pi
}{N}\right) ^{1-\beta }\right] ^{1/2} \cos \left( \frac{2\pi
ik}{N}+\phi _{k}\right), \label{generation}
\end{equation}
where $\phi _{k}$ are $N/2$ random phases uniformly distributed in
the interval $[0,2\pi]$ \cite{mou98,osb89,gre91}. Thus by
construction the power spectrum of the series $\{ \xi_i \}$ is of
the type $1/k^{\beta}$. By choosing different values for the
exponent $\beta$ we generate series of site energies with
different degree of spatial correlations: for $\beta =0$ we have
pure disorder (white noise), $\beta < 0$ corresponds to
anticorrelations, and $\beta > 0$ represents positive correlations
in the series of site energies $\{ \xi_i \}$. In our study we
consider only systems with positive correlations ($\beta \geq 0$).

Once the series $\{ \xi_i \}$ is obtained, we normalize it to zero
mean and unit standard deviation, thus fixing the width of the
site energy distribution to unity. This is equivalent to keeping
the `traditional disorder' of the system fixed, since the standard
deviation of the distribution of $\{ \xi_i \}$ quantifies the
variety in the site energies of the atoms forming the chain, while
their spatial order is quantified by the exponent $\beta$. Note
that for uncorrelated-disordered systems ($\beta=0$) the disorder
is quantified by the standard deviation in the case of site
energies randomly drawn from a Gaussian distribution, or by the
width of the box in the case of a box distribution.

After the normalization of the site energies, we diagonalize the
hamiltonian (\ref{hamil}) to obtain the energy spectrum $\{E_i\}$,
where $E_1<E_2<\ldots<E_N$. For any system size $N$ and any value of
the correlation exponent $\beta$ in our numerical
calculations, we diagonalize $2^{24}/N$ realizations of the
hamiltonian (\ref{hamil}). Thus we have a sufficiently large ensemble of
realizations to avoid statistical fluctuations in our results,
while we consider the same number of $2^{24}$ energy levels
for any $N$.

Once the energy spectrum is obtained, we study the
distribution of the spacings between consecutive energy
levels. Since the density of energy levels is not constant
throughout the energy band,
and thus the local average energy spacing is not constant either,
one cannot compare
fluctuations in the spacings obtained from different
regions of the band. To avoid this problem, we normalize to unity
the local average energy spacing from different regions of the energy band,
thus effectively normalizing all energy spacings to the same scale.
This `unfolding' of the energy spectrum is a procedure commonly used in the study of level
statistics of disordered systems \cite{boh84}. In brief the
unfolding procedure consists of the following steps: we first
introduce the integrated density of energy  levels $g(E)$ defined
as
\begin{equation}
g(E_i)=i. \label{int}
\end{equation}
Thus, $g(E_i)$ is the number of energy levels below the energy
$E_i$. Second, we fit $g(E)$ using a polynomial function. This fit
represents the averaged integrated density of energy levels
$\overline{g}(E)$. Next, the unfolded energy spectrum
$\{\varepsilon_i \}$ is obtained from the map
\begin{equation}
\varepsilon_i= \overline{g}(E_i). \label{map}
\end{equation}

To avoid unfolding problems related to irregular behavior of
$g(E)$ at the borders of the energy band, we consider only energy
levels from the central region of the band. Specifically for a
system of size $N$, we obtain $N$ energy levels (eigenvalues of
the hamiltonian (1)), and we consider the central part of the
spectrum $\{E_i\}$, where $i \in [N/3 +1,2N/3]$. We obtain the averaged integrated
density of levels $\overline{g}(E)$ by fitting $g(E)$ with a cubic
polynomial in the interval $[E_{N/3 +1},E_{2N/3}]$.

Using the unfolded spectrum $\{\varepsilon_i\}$, we study
the normalized distribution function of energy spacings
$P(s)$, where $s_i\equiv\varepsilon_{i+1}-\varepsilon_i$.
From Eqs. (\ref{int}), (\ref{map}), we have that
the average level spacing is $\left\langle {s} \right\rangle=1$.

\begin{figure}
\epsfig{file=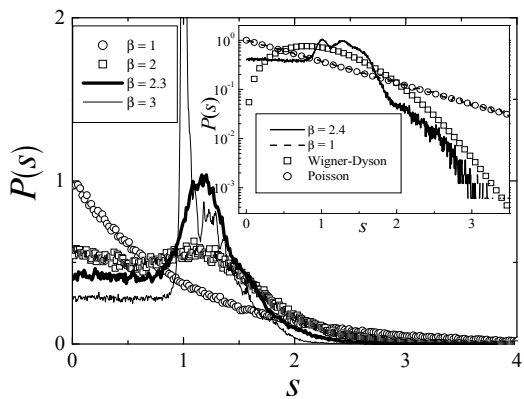,clip=,width=7cm}
\caption{Distribution functions $P(s)$ obtained for different values
of $\beta$ and for $N=2^{17}$. Inset: log-linear plot
of $P(s)$ for two values of $\beta$ and for Poisson and Wigner-Dyson (WD)
distributions. For $\beta =1 < \beta_{\rm c}$ $P(s)$ is Poissonian, while
for $\beta = 2.4 >\beta_{\rm c}$
 the tail of $P(s)$ decays even faster than for WD.}
\label{distributions}
\end{figure}

For the classical cases of 1D and 2D uncorrelated disordered
systems in the limit of large
system size $P(s)$ follows the Poisson distribution
\begin{equation}
P_{\rm P}(s)=e^{-s} \label{pois}.
\end{equation}
A Poisson distribution for $P(s)$ indicates strong clustering
between energy levels because it reaches maximum when $s
\rightarrow  0$ (Fig.~\ref{distributions}).

\begin{figure}
\epsfig{file=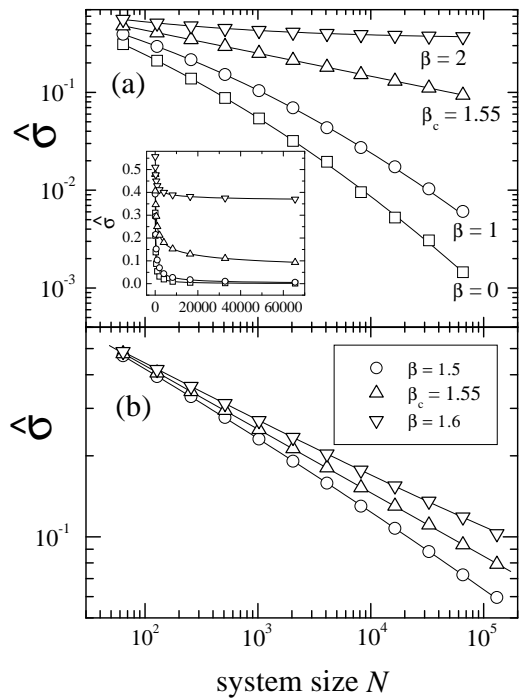,clip=,width=7cm}
\caption{Log-log plot of $\hat \sigma$ as
a function of the system
size $N$ for different values of the correlation exponent $\beta$.
Inset: same dependence in linear scales. Solid lines represent
fits with Eqs. (\ref{below}), (\ref{at}) and
(\ref{above}).}
\label{hatsigma}
\end{figure}

In our analysis we characterize $P(s)$ using its standard
deviation $\sigma$. For the classical case of Poissonian form for
$P(s)$ (\ref{pois}), we have $\sigma\equiv\sigma_{\rm P}=1$.
For convenience, we study the behavior of $\hat \sigma\equiv
1-\sigma$. In Fig.~{\ref{hatsigma}} we show  $\hat \sigma$
as a function of the system size $N$ for
different values of the correlation exponent $\beta$.
For uncorrelated disorder ($\beta=0$) and for large $N$,
$\hat \sigma \rightarrow 0$ (or equivalently $\sigma \rightarrow 1$),
indicating Poissonian behavior for $P(s)$ as expected for 1D disordered
systems \cite{wis90}.

\begin{figure}
\epsfig{file=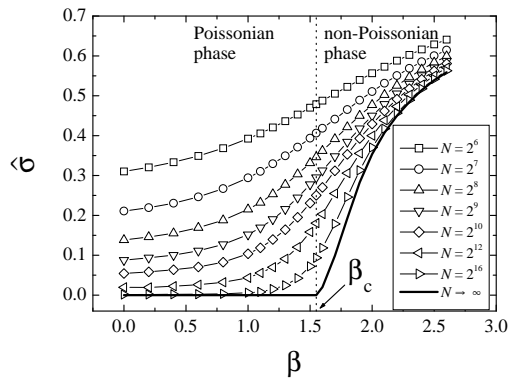,clip=,width=7cm} \caption{Dependence of
$\hat \sigma$ on the correlation exponent $\beta$ for varied
system size $N$. The solid line represents the behavior of $\hat
\sigma$ in the limit of $N\rightarrow \infty$ (i.e., $\hat \sigma_{\infty}$)--- a phase
transition from Poissonian ($\hat \sigma = 0$) to non-Poissonian
($\hat \sigma \neq 0$) level statistics at the critical value
$\beta_c$.}
\label{sigmainf}
\end{figure}

Introducing certain degree of spatial correlations ($\beta >0$) in
the system we find that in the limit of large $N$ the level
statistics exhibits again Poissonian behavior ($\hat \sigma
\rightarrow 0$, Fig.~\ref{hatsigma}a), indicating energy level clustering and
electronic localization. This finding indicates that Poissonian
behavior in the level statistics exists even in the presence of
long-range correlations. However, we find that the convergence of
$P(s)$ to the Poisson distribution with increasing $N$ is much
slower in the case of correlated disorder ($\beta>0$) compared to
the case of uncorrelated disorder ($\beta=0$)(Figs.~\ref{hatsigma} and \ref{sigmainf}).

Increasing the strength of the long-range correlations we find a threshold value,
a critical exponent $\beta_c$, above which the functional form of $P(s)$
does not converge to Poisson distribution --- i.e., $\hat \sigma$ does not converge to zero
in the limit of large $N$  (Figs.~\ref{hatsigma}a and \ref{sigmainf}). At the critical value $\beta_c$ we find
that for increasing $N$
the level statistics converges very slowly to Poisson distribution
and $\hat \sigma \rightarrow 0$  as a power law (Fig.~\ref{hatsigma}b).

We next investigate the functional dependence of $\hat \sigma$ on
the system size $N$ for the three regimes: (i)  $\beta < \beta_c$;
(ii) $\beta =\beta_c$ and (iii) $\beta > \beta_c$. Based on
simulations of various systems sizes up to $N=2^{17}$ we model the
behavior of $\hat \sigma$ using the following expressions:
\begin{eqnarray}
\hat \sigma &=& a_1 N^{-b_1-c_1 \log N}  \quad (\beta < \beta_{\rm c}) \label{below}\\
\hat \sigma &=& a_2 N^{-b_2}  \quad (\beta = \beta_{\rm c}) \label{at}\\
\hat \sigma &=& \hat \sigma_{\infty}+a_3 N^{-b_3}  \quad (\beta >
\beta_{\rm c}) \label{above}
\end{eqnarray}
where all the parameters are positive and in general depend on the
correlation exponent $\beta$. To test the validity of these
expressions, we use the Levenberg-Mardquardt (L-M) algorithm
(\cite{NumRec}) to fit the data in Fig.~\ref{hatsigma} a,b and to estimate the
optimal values of the parameters. Although we have tried more expressions to fit the data,
those in Eqs. (\ref{below}), (\ref{at}) and (\ref{above}) are the best we have found attending two criteria:
good description of data, and less number of parameters.

We note that for $\beta < \beta_{\rm c}$ the behavior of $\hat
\sigma$ vs. $N$ on log-log plot presents negative curvature
(Fig.~\ref{hatsigma}a), while for $\beta=\beta_{\rm c}$, $\hat \sigma$ depends
on $N$ as a power-law with negative slope, so that in both cases
for $N~\rightarrow~\infty$, $\hat \sigma~\rightarrow~0$ (Fig.~\ref{hatsigma}b)
indicating that in the thermodynamic limit the level statistics is
Poissonian. In contrast, for $\beta > \beta_{\rm c}$, the
dependence of $\hat \sigma$ on $N$ has positive curvature
(Fig.\ref{hatsigma}b), indicating a decay slower than a power-law.
We find that data are best modeled (with best fit based on the L-M
algorithm) as a power law with an additive positive constant $\hat
\sigma_{\infty}$ such that $\hat \sigma=\hat \sigma_{\infty}$ for
$N \rightarrow \infty$ (\ref{above}). Thus our results suggest
that for $\beta > \beta_{\rm c}$, in the thermodynamic limit $\hat
\sigma$ does not converge to zero, and that the level statistics
is not of Poissonian type.

To determine $\beta_c$ we use the following procedure: (i) starting from small values of $\beta$
for which the dependence of $\hat \sigma$ on $N$ follows Eq. (\ref{below}),
we increase $\beta$ and observe that the fitting parameter $c_1$ decreases,
and for a given value of $\beta$ becomes zero, so that Eq. (\ref{below}) is not
valid anymore; (ii) starting from large values of $\beta$ for which
$\hat \sigma$ follows Eq. (\ref{above}), we decrease $\beta$
and observe that the fitting parameter $\hat \sigma_{\infty}$ decreases, and for a given value of $\beta$ becomes
zero, so that Eq. (\ref{above}) is not valid anymore. We find that for both (i) and (ii) the
transitions occur at a critical value of $\beta\equiv \beta_c=1.55 \pm 0.05$ (Figs. \ref{hatsigma} and \ref{sigmainf}),
where the behavior of $\hat \sigma $ is described by Eq. (\ref{at}).

We next obtain a phase diagram of the properties of the level
statistics as a function of the degree of spatial correlations in
the system. We systematically investigate the asymptotic behavior
of $\hat \sigma$ in the limit of large system size $N$ as a
function of the correlation exponent $\beta$ (Fig. \ref{sigmainf}). For a fixed
system size $N$ we calculate how $\hat \sigma$ depends on the spatial correlations choosing a dense set of $\beta$ values. We
then repeat the calculations for increasing $N$. We find  that
$\hat \sigma$ is an increasing function of $\beta$, and that for
each value of $N$ a relatively flat region at small $\beta$ is
followed by a sharp increasing in $\hat \sigma$ for large $\beta$
(Fig.~\ref{sigmainf}). This behavior becomes  more pronounced with increasing
$N$. Further, we find that the flat region in $\hat \sigma$
extends to intermediate values of $\beta$ and rapidly approaches
zero with increasing $N$. This is in agreement with our finding of
level statistics of Poissonian type even in the presence of a
moderate degree of spatial correlations in the system, and with
the predictions of Eqs. (\ref{below}),(\ref{at}). In contrast, for
large values of $\beta$, the values of $\hat \sigma$ remain large
and do not decrease substantially with increasing the system size
$N$. Thus, we observe a transition in $\hat \sigma$ centered at
intermediate values of $\beta$, which becomes more abrupt with
increasing $N$. To extrapolate the behavior of the level
statistics in the thermodynamic limit and for large $\beta$, we
use Eq. (\ref{above}) since for $N \rightarrow \infty$, $\hat
\sigma \rightarrow \hat \sigma_{\infty}$. We estimate $\hat
\sigma_{\infty}$ for a dense set of large and decreasing values of
$\beta$--- the solid thick line in Fig. \ref{sigmainf}, which sharply decreases to $\hat \sigma _{\infty}= 0$
for $\beta \equiv \beta_c=1.55$. This suggests a phase transition from
a Poissonian behavior of the level statics characterized by $\hat \sigma_{\infty}=0$ for $\beta < \beta_c$,
indicating strong clustering between energy levels, to a non-Poissonian phase defined by
$\hat \sigma_{\infty} \neq 0$ for $\beta > \beta_c$. As $\hat \sigma_{\infty}$ is a function of $\beta$,
this suggests that for any $\beta > \beta_c$ a different level statistics $P(s)$ is obtained.
Thus we find a new class of correlated
disordered systems characterized by energy level repulsion, different values of
$\hat \sigma_{\infty}$ and different distribution functions for the energy spacings $P(s)$.

\begin{figure}
\epsfig{file=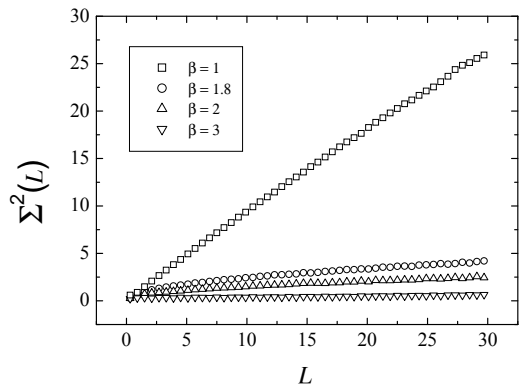,clip=,width=7cm}
\caption{Behavior of $\Sigma^2$ as a function of $L$ for spectra obtained
for different values of $\beta$ and for a system size $N=2^{17}$.}
\label{sigma2}
\end{figure}

Similar conclusions can be drawn with the study of $\Sigma^2(L)$, i.e., the variance
of the number of levels in boxes of length $L$ in the unfolded spectrum. It is known that for
Poissonian behavior, $\Sigma^2(L)$ is linear with $L$ with slope 1. We obtain this linear behavior
for any $\beta < \beta_c$ (see the case $\beta=1$ in Fig.\ref{sigma2}), indicating Poissonian behavior,
in agreement with our previous results. For $\beta > \beta_c$, we obtain a non-linear and slow increasing of $\Sigma^2(L)$
as a function of $L$ (slower for increasing $\beta$), indicating level repulsion
and non-Poissonian behavior, also in agreement with our previous results, and with
the behavior of the $P(s)$ functions (see below). 

In Fig.~\ref{distributions} we show  $P(s)$ for several values of
$\beta$ and for finite system
size $N=2^{16}$. We note that in the thermodynamic limit of large $N$ the form of
$P(s)$ may change. However, this change is expected to be not
significant, since for $N=2^{16}$, $\hat \sigma~-~\hat
\sigma_{\infty}~\simeq ~0.005$.
In general, we obtain that for $\beta < \beta_c$, although there exists a moderate degree
of correlations in the system,
$P(s)$ is exponential (Poissonian) (see the case $\beta=1$ in Fig.\ref{distributions}).
When $\beta > \beta_c$, the Poissonian behavior is lost: as $\beta$ departs from
$\beta_c$, the functions $P(s)$ for low $s$ decrease gradually, and
simultaneuously, an increasing peak for increasing $\beta$ appears 
at $s=\langle s \rangle~=1$, indicating strong level repulsion. For extreme
values of $\beta$, $P(s)$ for low $s$ is very small, while the peak at $s=1$
becomes huge (see the case of $\beta=3$ in Fig.\ref{distributions}).
This finding is consistent with the expectation that
extreme values of $\beta$ correspond to an ordered system, for
which the level statistics is of the type $P(s)=\delta(s-1)$.

In summary we find that introducing spatial long-range
correlations in 1D disordered systems leads to a transition from
a Poissonian to a new class of functional forms describing the level
statistics in the thermodynamic limit. Further, we find a critical
value for the correlations below which the level statistics
exhibit Poissonian behavior associated with energy level
clustering, similar to the one observed in uncorrelated disordered
systems. Above this critical value the system is characterized by
level repulsion. These findings may relate to previous reports on
localization-delocalization transition in the electronic
properties of 1D systems driven by spatial correlations in the
disorder \cite{mou98}. In that work, the transition is detected at $\beta=2>\beta_c$,
where we already observe non-Poissonian behavior, as expected in the
extended regime. 

\acknowledgments We thank the Spanish Ministerio de Ciencia y
Tecnolog\'{\i}a for support (grant number BFM2002-00183).

\end{document}